# Electrical processes in planetary atmospheres

Karen Aplin, Martin Airey*, Elliot Warriner-Bacon
Physics Department
University of Oxford, UK
phone: (44) 1865-273491
e-mail: karen.aplin@physics.ox.ac.uk

*also at Department of Meteorology, University of Reading, UK

***Abstract*—Lightning is common throughout the Solar System, and charging of particles occurs in all atmospheres due to ionization from galactic cosmic rays. Here, some electrical processes relevant to the atmosphere of Venus are outlined and discussed in a comparative planetology context.**

## I. Introduction

Observations of lightning at distant planets shows that some atmospheres have the right meteorological conditions to separate charge in their clouds. Lightning-induced chemical effects could be relevant for the origin of life, and lightning is also helpful in providing a marker of strong atmospheric motion [1]. In addition to charge separation leading to electrical discharges, ionization by cosmic rays provides a charge source in all atmospheres, which may be augmented by other processes, for example, natural radioactivity [2]. Ions and electrons can influence clouds and hazes; this is particularly important in atmospheres distant from the Sun such as Neptune and Pluto, where there can be little other energy input [3]. Lightning or other charge separation mechanisms can combine with atmospheric ionization to allow atmospheric current flow, and in some cases to form a global electric circuit [1]. The consequence of a global electric circuit (GEC) is the transfer of charge throughout an atmosphere, connecting lightning-generating regions with charge-sensitive processes at distance.

Atmospheric electricity on Venus is a particularly interesting and timely example, with the realistic possibility of lightning, cloud charging effects and a global electric circuit. The existence of lightning has long been debated but recent observations do seem to indicate its presence, although there has been no conclusive optical detection [e.g. 4, 5]. The Lightning and Airglow Camera (LAC) on the Japanese Akatsuki mission is currently orbiting Venus, providing the best opportunity yet for optical detection, despite the limited flyby opportunities after Akatuski failed to achieve its planned orbit insertion in 2010. The rest of this paper will consider atmospheric electrical processes which may be present on Venus, with a particular focus on laboratory analogue experiments.

## II. Lightning on Venus

Venus' atmosphere is a thick layer of carbon dioxide with permanent and complete cloud cover in a dense deck from ~48 to 70 km altitude. Lightning in this environment is expected to occur either

1. within the clouds (i.e. intra-cloud lightning)
2. below the clouds (i.e. cloud-to-ground lightning)
3. upward from the clouds (i.e. transient luminous events).

Any cloud-related lightning requires an efficient mechanism both to charge droplets and particles in the clouds, and then to separate the charge enough for the breakdown potential to be achieved. On other planets with lightning, such as Jupiter and Saturn, the discharges are thought to originate from atmospheric regions where there are clouds containing liquid water and ice, together with strong convection, allowing the terrestrial model of thunder-cloud charging to be assumed [1]. However, this assumption cannot be made for Venus; the clouds there are layer clouds, not strongly convective, and the sulfuric acid does not exist in ice form, which would be required for a terrestrial-like mechanism. This does not rule out other, as yet unknown, charging mechanisms [1]. Another problem is that models suggest that the cloud environment is too conductive for effective charge separation [6], which excludes type (1) above based on our current understanding. According to [7] and [8], Venus' atmosphere may support upward-directed lightning such as sprites, and these discharges should be suitable for optical detection by Akatsuki. However, since sprites on earth are always associated with thunderstorms, it is not clear if they could exist if the clouds cannot support charge separation, i.e. lightning type (3) may be dependent on the existence of lightning type (1).

Option (2), lightning below the clouds, would be consistent with lack of optical observations of lightning, since the optical emissions would be obscured by the thick cloud layer. Cloud-to-ground lightning would however require a significant charge in the base of the cloud. Although the pressure at cloud level is similar to that at the surface of the Earth, implying a similar breakdown voltage to that on Earth, the discharge would have to overcome both the 50 km to the surface, and the increase with breakdown voltage with pressure up to 100 MV $m^{-1}$ at the surface. Cloud-to-ground lightning on Venus therefore seems highly unlikely.

Almost the only feasible remaining possibility is volcanic lightning. There is evidence that Venus may have active volcanoes that erupt explosively, which could generate volcanic lightning similar to on Earth [9]. Fractoemission, the release of charge through breaking rocks, generates charged ash particles near the vent of a volcano, but its efficiency for an extra-terrestrial atmosphere has never been investigated. The next section describes lab experiments studying fractoemission in a high-pressure carbon dioxide atmosphere similar to Venus.

## III. Laboratory Study of Fractoemission

An experimental system was developed at Oxford to simulate conditions similar to those that occur on Venus. The facility can simulate the high-pressure, $CO_2$-dominated atmosphere of Venus at ~10 km altitude (~5 MPa). A key finding of previous work [8] is that ash plume-forming eruptions (potentially capable of producing electric fields and lightning within the ash column) are much more likely to occur at higher altitudes such as these on Venus. The high temperature of Venus' atmosphere was not simulated in this pilot study.

The components of the 1-litre chamber include temperature/pressure monitoring and logging equipment, an embedded rock collision apparatus to generate the charged rock fragments, and a copper collection electrode connected to a sensitive electrometer. The rock fragmentation mechanism comprises a solenoid upon which a rock sample is attached and, when driven, collides with a second, fixed, rock sample. The repeated impacts effectively generate charged rock fragments through fractoemission that subsequently land on a copper Faraday plate. The apparatus is controlled and monitored using the LabView software package. Figure 1 shows the simulation tank and connections.

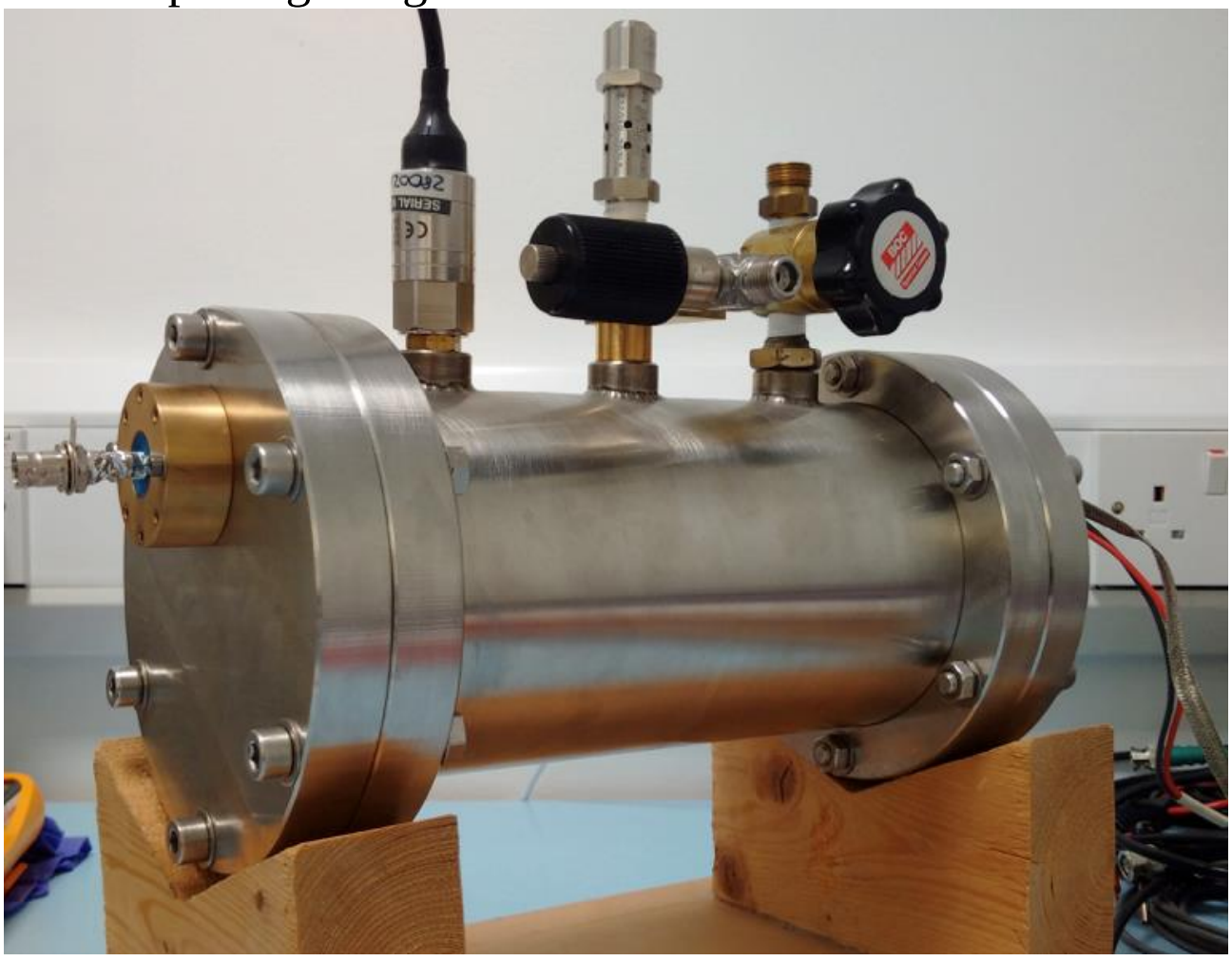

Fig. 1. Photograph of the high-pressure atmospheric simulation chamber used in the ash-charging experiments. The electrical connection to the electrometer can be seen on the left hand side, gas valves on the top and the power supply wires to the solenoid are emerging from the far end.

A series of experiments was devised to systematically characterize the effects of pressure and atmospheric composition. Pressure was varied over the range ~600 Pa to ~4 MPa in air and $CO_2$ atmospheres, and composition was varied from ambient air, artificial (dry) air to pure $CO_2$ at constant pressures in order to isolate the effects of the individual variables. Secondary experiments also included varying the oxygen content of air (~21% to <1%).

From these experiments, it was found that all ash charging by fractoemission resulted in net negatively charged ash particles, as a negative current was detected on the Faraday plate during every run producing ash particles. In ambient pressure air, the currents produced due to fractoemission were ~ -0.2 pA. The effect of increasing the pressure was to decrease the mean free path resulting in decreased ion mobility, therefore reducing the ash charging. The low-pressure experiments had the opposite effect, enhancing currents to ~ -0.5 pA in air at 600 Pa.

Decreasing the partial pressure of $O_2$ (or increasing $N_2$) in air also cause an increased current. This is thought to be due to a decrease in concentration of species with the greater electron affinity ($O_2$) which would normally rapidly attach to any free electrons created by fractoemission. At lesser $O_2$ concentrations, free electrons are available for collision with ash particles for longer, to increase the negative charge available on the ash. Replacing air with $CO_2$ also increased the ash charging, with 95% $CO_2$ resulting in greatly enhanced currents of >5 pA. This, again, is thought to be due in part to the difference in electron affinity; $CO_2$ does not form stable anions, with the $CO_2^-$ ion decaying to release a mobile electron after ~100μs, which then becomes available again to charge ash particles [10].

Figure 2 plots the ratio of currents produced in a high-pressure, $CO_2$ simulation to those produced in ambient pressure air (effectively a Venus to Earth ratio in terms of pressure and composition). This range of pressures is consistent with regions on Venus theoretically capable of hosting explosive volcanic plumes (down to ~13 km above Venus' mean planetary radius). The overall effect of a high-pressure $CO_2$ 'Venus plume environment' is an increase in fracto-current at pressures up to 36 bar (3.6 MPa), which is up to five times more efficient than terrestrial fracto-charging. If the decrease in fracto-current at pressures greater than 36 bar is robust, it suggests that the $CO_2$ effect to enhance charging dominates over the compensating pressure effect. More measurements at higher pressure may result in a shift to the pressure effect dominating, leading to a reduction in current enhancement. Future work will clarify these findings.

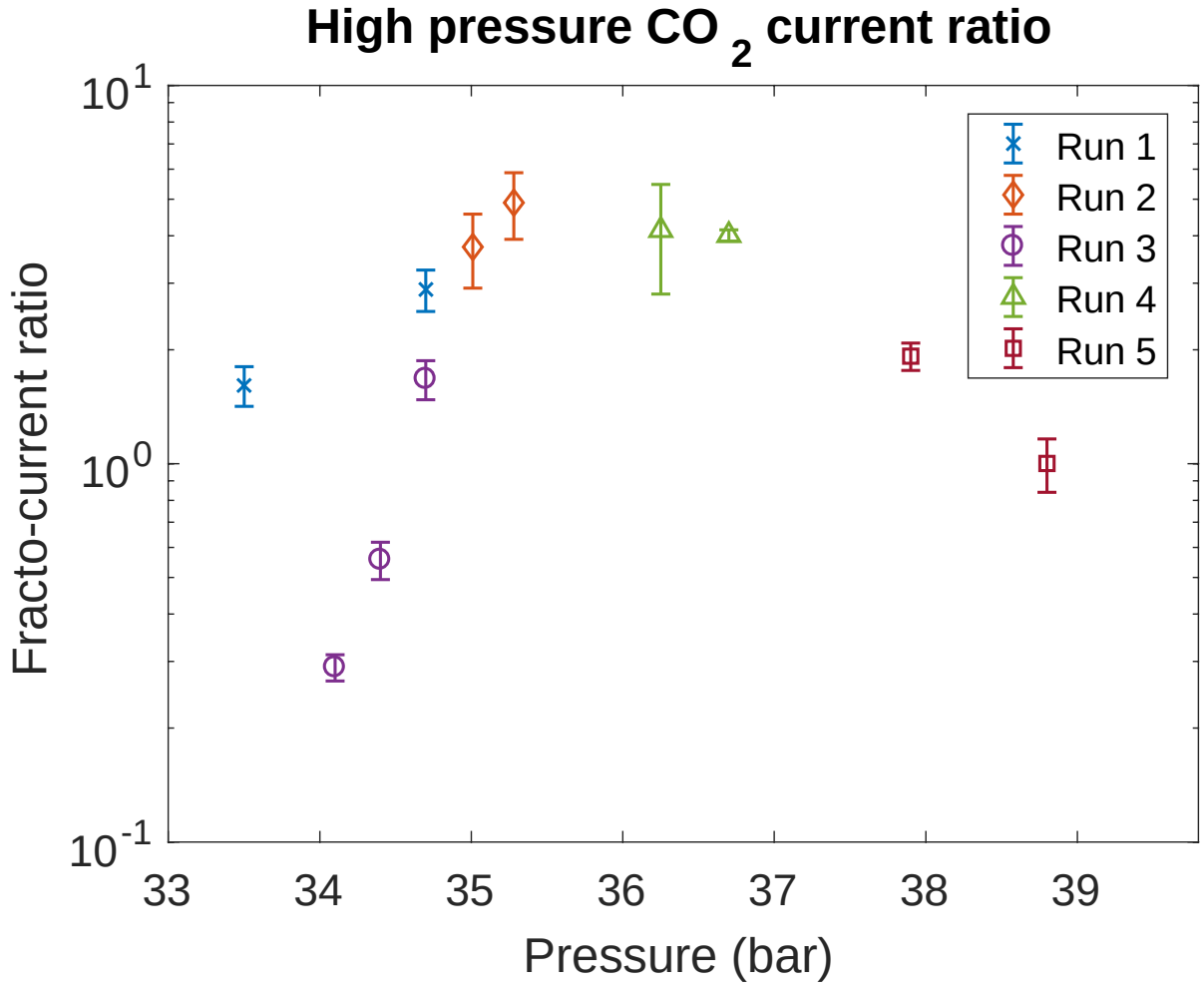

Fig. 2. Plot of high-pressure $CO_2$ to ambient air fracto-current ratio over a range of pressures consistent with regions on Venus capable of hosting explosive volcanic plumes [8]. The points represent repeated experiments with the same composition.

## IV. FAIR WEATHER PROCESSES

In addition to lightning, atmospheric electrical processes may affect the cloud behavior on Venus, through ionization from galactic cosmic rays (GCRs). The cloud deck comprises three distinct layers made up of characteristic droplet sizes, generally small in the upper cloud layer (~1 μm) and larger in the middle and lower cloud layers (~1.4 to 3.6 μm), with very small particles (<1 μm) making up haze layers above and below as well as persisting throughout the whole cloud deck [11]. The atmospheric and particle compositions play an important role in determining the electrical structure and particle size and charge distribution within the clouds. Understanding the microphysical and electrical structure of this complex environment is important in assessing the dynamic role in atmospheric structure, precipitation, possible current flow in a global electric circuit (GEC), and hazards to future probes. This knowledge is also important in order to effectively assess the likelihood and occurrence of electrical discharges in this environment.

### *A. Cloud charging*

One consequence of charging is a change in the critical minimum supersaturation at which droplets begin to form [12]. Figure 3 shows the Köhler curves (which, at their maxima, show the size and supersaturation at which droplets become stable) for droplets in Venus' clouds.

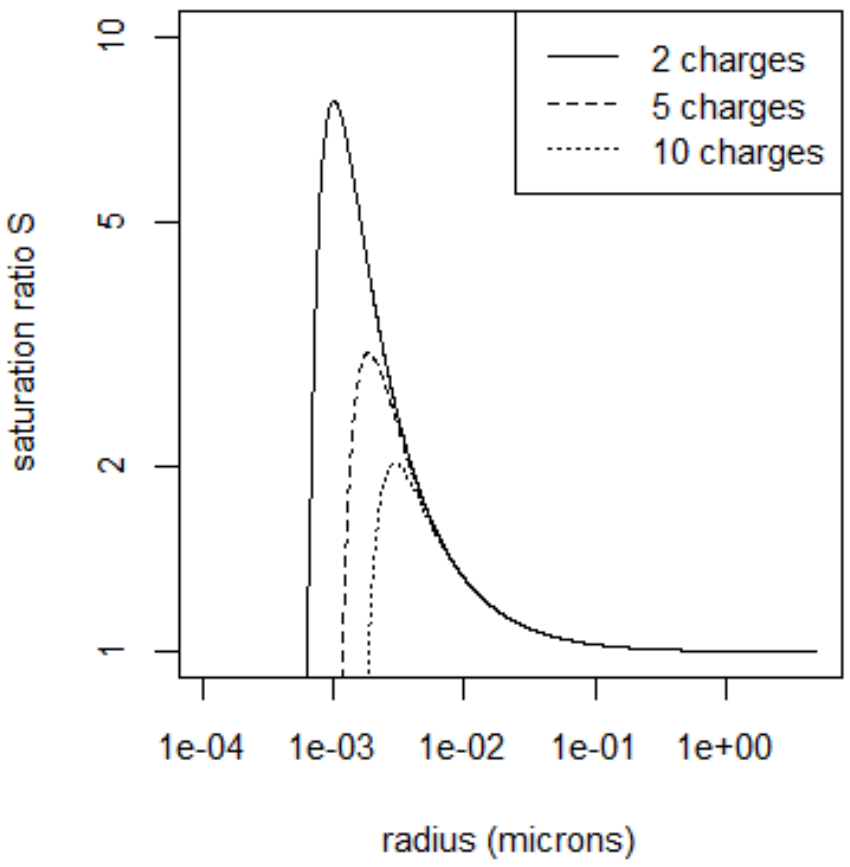

Fig. 3. Variation of saturation ratio S with droplet radius, for increasing numbers of elementary charges added to the droplet, Venus-like conditions in the cloud deck. (The critical supersaturation is given by the turning point of each plotted line.) These high saturation ratios are possible in Venus's atmosphere.

The charging acts to stabilize droplets against evaporation, thereby allowing slightly larger drops to exist at lower saturation ratios. A major source of charge in planetary atmospheres is from cosmic ray ionization, which has recently been demonstrated to influence clouds in the highly-supersaturated atmosphere of Neptune [3]. The dense sulfuric acid atmosphere of Venus provides another set of favorable circumstances for these effects. Condensation of gas directly onto atmospheric ions is a potential source of fresh droplets on Venus because supersaturation is common there, whereas this process would be impossible on Earth [2]. Highly charged particles allow condensation at lower temperatures than neutral particles, and this effect is modelled using Köhler theory; it could therefore be an important contributor in determining cloud droplet nucleation altitude.

On Earth, layer clouds become charged at the upper and lower horizontal boundaries due to a sharp change in conductivity at the cloud-air transition; the current travelling through the cloud, as a component of the GEC, results in the accumulation of charge at this transition [12, 13]. In addition to this, the droplets also charge and discharge by diffusion of cluster ions. By assessing the extent to which these processes occur in the Venusian atmosphere, a model for understanding haze formation and broader cloud behavior emerges. The findings also provide a means to quantify any properties of a possible Venusian GEC, which, if it exists, could act to distribute charge more completely through the Venusian atmosphere, to further influence clouds.

### *B. The role of galactic cosmic rays*

GCRs are important for ionization in atmospheres across the Solar System. In Earth's atmosphere, cluster-ions created by GCRs play a role in meteorological processes [14]. Venus is particularly likely to be affected by GCR-cloud interactions, because the maximum in cosmic ray ionization at ~60 km occurs at the same height as the main cloud deck [15], so the clouds are being directly and continuously injected with charge, separated in turn by any local electrical fields. Further, the relative proximity of Venus to the Sun means that its cloud-producing altitudes are disproportionately affected by space weather events, with a likely increase in ionization rate of several orders of magnitude from solar energetic particles [15]. This presents a contrasting aspect for comparative planetology studies, since on Earth the major cloud-forming altitudes are in the lower troposphere, well below the GCR maximum height and therefore beyond all but exceptional space weather events. There is a need to understand ionization effects on charging of Venus-like cloud droplets. This could be carried out by laboratory analogue experiments using an electrostatic levitator to apply charge to well-characterized sulfuric acid droplets and measure its effects on growth and evaporation. The enhanced ionization effects during solar storms could also be simulated in this way.

## V. CONCLUSION

The atmospheric electrical environment of Venus is currently being studied both by the Akatsuki space mission, and in ongoing and planned laboratory analogue experiments. Preliminary results investigating fractoemission efficiency in a Venus analogue atmosphere indicate that fracto-charging reduces because of the increased pressure, as expected, but that the carbon dioxide atmosphere is more likely to support the existence of free electrons, which has the overall effect of enhancing ash charging at pressures up to 3.6 MPa. Further work is need both to confirm this effect and to determine whether the enhanced fracto-charging could generate volcanic lightning on Venus. This would be achieved by measuring charge separation, and also by heating the tank to provide a more realistic temperature regime. Experiments are also planned to study the charging of Venus cloud droplets, to understand the effects of cosmic rays and space weather events on cloud microphysics.

The Venus example discussed here is an interesting study of the similarities and differences in atmospheric electrical processes compared to Earth. In some cases, known physical effects, such as the direct (Wilson) condensation of gases onto ions, may take place despite being impossible on Earth, whereas other processes, such as charging at cloud edges, may be similar. Laboratory experiments under Venus-like conditions have demonstrated how poorly understood fractoemission is, but that it may be more efficient in other planetary environments. Looking to the future, the many exoplanets now known to orbit other stars is likely to further broaden the range of known atmospheric electrical processes, though their detection is expected to be challenging.

We acknowledge helpful discussions with Prof Y. Takahashi and others on the Akatsuki LAC science team.